\documentclass[
aps,
prl,
longbibliography,
superscriptaddress,
 amsmath,amssymb,
twocolumn,
]{revtex4-1}

\usepackage{graphicx}
\usepackage{dcolumn}
\usepackage{bm}
\usepackage{upgreek}
\usepackage{color}
\usepackage{hyperref}
\DeclareMathAlphabet{\mathpzc}{OT1}{pzc}{m}{it}

\newcommand{\eref}[1]{Eq.~(\ref{#1})}
\newcommand{\fref}[1]{Fig.~\ref{#1}}
\newcommand{\puoli}{\frac{1}{2}}

\begin{document}

\title{Landau-Zener-St\"uckelberg interference in a multimode  electromechanical system in the quantum regime}

\author{Mikael Kervinen}
\affiliation{Department of Applied Physics, Aalto University, P.O. Box 15100, FI-00076 AALTO, Finland}
\author{Jhon E. Ram\'{i}rez-Mu\~noz}
\affiliation{Departamento de F\'{i}sica, Universidad Nacional de Colombia, 111321 Bogot\'{a}, Colombia}
\author{Alpo V\"alimaa}
\affiliation{Department of Applied Physics, Aalto University, P.O. Box 15100, FI-00076 AALTO, Finland}
\author{Mika A. Sillanp\"a\"a}
 \email{Mika.Sillanpaa@aalto.fi}
\affiliation{Department of Applied Physics, Aalto University, P.O. Box 15100, FI-00076 AALTO, Finland}%



\begin{abstract}
The studies of mechanical resonators in the quantum regime not only provide insight into the fundamental nature of quantum mechanics of massive objects, but also introduce promising platforms for novel hybrid quantum technologies. Here we demonstrate a configurable interaction between a superconducting qubit and many acoustic modes in the quantum regime. Specifically, we show how consecutive Landau-Zener-St\"uckelberg (LZS) tunneling type of transitions, which take place when a system is tuned through an avoided crossing of the coupled energy levels, interfere in a multimode system. The work progresses experimental LZS interference to cover situations where the coupled levels are those of a qubit interacting with a multitude of mechanical oscillators in the quantum limit. The work opens up applications in controlling multiple acoustic modes via parametric modulation.
\end{abstract}

\maketitle

Advances in the control over mechanical degrees of freedom have taken great leaps forward allowing to engineer experiments that go deep into the quantum regime, consequently showing the underlying nature of the quantized vibration energy \cite{OConnell2010,Chu2018,Satzinger2018,Viennot2018,Lehnert2019Fock,Safavi2019Fock,Cleland2019PhEntangl}. These works predominantly utilized superconducting quantum bits combined with a variety of different types of mechanical resonators that can be accessed resonantly through the qubit in the high gigahertz frequency range. The resonators can be made with surface acoustic waves (SAW) \cite{Delsing2014,Nakamura2017SAW,Leek2017SAW,Astafiev2018SAW,Lehnert2019Fock,Cleland2019PhEntangl,Delsing2019SAW}, phononic crystals \cite{Safavi2019Fock}, or high overtone bulk acoustic wave resonators (HBAR) \cite{SchoelkopfHBAR2017,Chu2018,Kervinen2018}, with piezoelectric materials allowing for strong coupling between electric and mechanical quantities. Mechanical modes are better isolated from the electromagnetic environment and can have longer coherence times than superconducting qubits. They are also much more compact than microwave cavities \cite{Naik2017RandomElectrodynamics,Sundaresan2015BeyondCavity}. Therefore, incorporating  mechanical resonators is highly appealing in quantum computing that can utilize harmonic oscillators \cite{Devoret2013CatCode,Devoret2015CatCode}.

In an HBAR system, the modes mostly reside in the substrate chip and hence feature diluted strain and low acoustic losses. The system exhibits a dense spectrum of acoustic modes that interact near resonance with the qubit, suggesting a possibility to manipulate the many-mode system through the qubit. One way to do the latter is to combine slow adiabatic changes and abrupt rotations of the adiabatic basis. This type of control of qubits resembles a coherent version of Landau-Zener tunneling transitions, which have been studied extensively in various two-level systems either in quantum or classical limit. These include superconducting qubits \cite{OliverLZ,LZ,Delsing2007LZ,Ilichev2008LZ,Oliver2008,Ashhab2010LZ,Han2010LZ,Han2011LZ,Petta2013LZtransmon,Li2013,Han2014LZgeom,Silveri2015StuckelbergModulation,Ilichev2016LZ,Silveri2017QuantumModulation}, nanomechanical systems \cite{LaHaye2009,LZbec2009,Pekola2011GLZ,Nori2012LZnems,Weig2016LZ,Muraki2018LZqd}, Bose-Einstein condensates \cite{LZbec2009,Nori2011LZbec,LZbec2014}, optical lattices \cite{Weitz2010OptLZ}, and other systems \cite{Grimm2007LZ,Petta2010LZ,Du2011nvLZ,Petta2012LZ,LZ2016Qdot}.

\begin{figure}[h]
  \begin{center}
    {\includegraphics[width=0.4\textwidth]{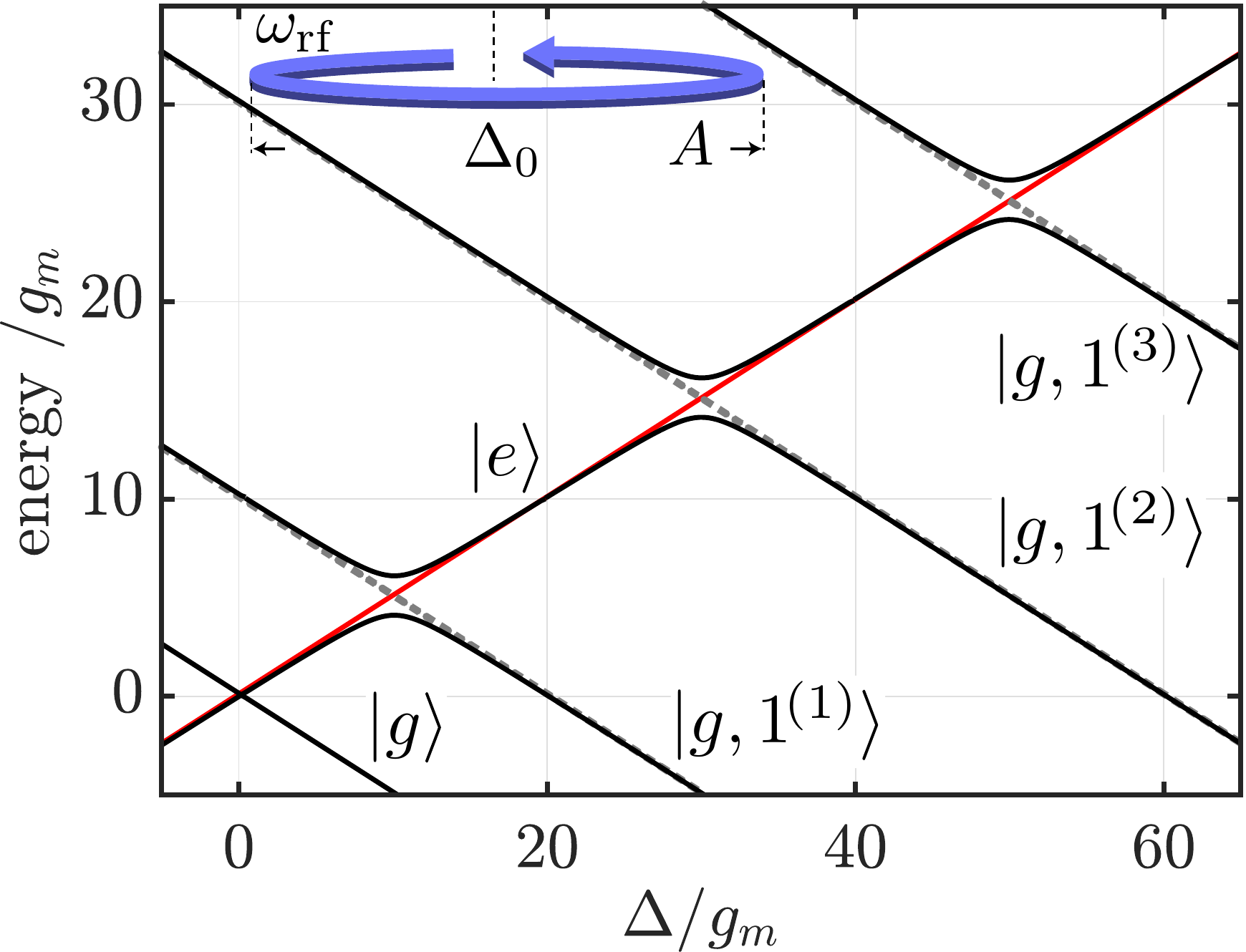}} 
    \caption{\emph{Photon-assisted Landau-Zener-St\"uckelberg  (LZS) interference in a multimode qubit-oscillator system.} The solid lines represent eigenvalues from \eref{eq:HLZ}. We restrict for simplicity to the lowest excitation manifold; for example, $|g, 1^{(3)}\rangle$ means that the qubit is in the ground state, and the harmonic mode number $3$ has one photon, and the rest of the oscillators are in the ground state. The arrow sketches a slow modulation of the bare qubit energy splitting represented by the blue and red lines. The dashed lines are the energies of three harmonic modes.}
    \label{fig:LZscheme}
 \end{center}
\end{figure} 

\begin{figure*}
  \begin{center}
   {\includegraphics[width=0.99\textwidth]{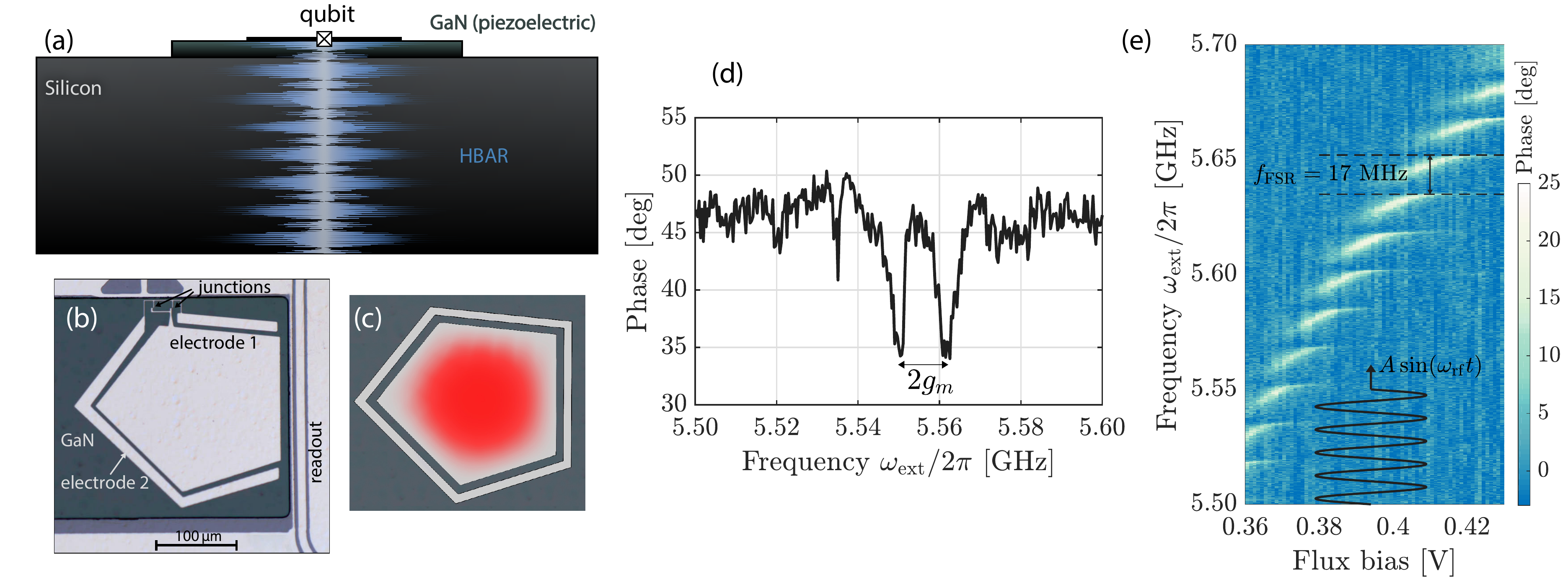}} 
    \caption{(a) Schematic cross-section of the high-overtone bulk acoustic (HBAR) modes that are located inside the massive substrate. 
    (b) Photograph of the device shows a transmon qubit that has an irregular pentagon shape which suppresses the laterally propagating modes. (c) Simulation of the mode displacement profile of one overtone acoustic mode. (d) Two-tone spectroscopy showing the vacuum Rabi splitting in the qubit on resonance with a mechanical mode number $i= 319$ at $\omega_0/2\pi \simeq 5.554$ GHz. (e) Spectroscopy as a function of flux bias, and a sketch of the slow bias modulation.}
    \label{fig:qubit}
 \end{center}
\end{figure*} 

In Landau-Zener-St\"uckelberg (LZS) interference, the system energy levels are modulated back and forth through an avoided crossing at a frequency $\omega_\text{rf}$ faster than the decay rates.
Earlier work on LZS physics in the quantum regime has focused on two-level systems, aside from theoretical considerations \cite{Brundobler_1993,Tatsuya1997,Sinitsyn2002multiLZ,Hanggi2008,Nori2012phLZ,Ashhab2014LZqed,Ashhab2017}.  Yet, systems possessing more complicated spectrum, if exposed to LZS conditions, can give new insight in manybody phenomena, or possible applications in quantum control. In the current work we address this situation. In addition to studying qubit-oscillator system in a regime where it can be treated as LZS interference, we concentrate on a truly multimode system, where the oscillators are acoustic modes in contrast to earlier work in the quantum regime.

Let us recall the case of a traditional LZS interference, as discussed in many works. We consider a quantum two-level system with the energy splitting $\omega_{\mathrm{0}}(\Phi)$. The splitting depends on a control parameter $\Phi$, which can be the flux through a SQUID loop as in this work. The levels are assumed to couple at the energy $\Omega$, resulting in the energies $\omega(\Phi) = \pm \sqrt{\omega_{\mathrm{0}}^2(\Phi) + \Omega^2}$ between the ground state and the excited state of the coupled system, and the avoided crossing equal to $2\Omega$ at the degeneracy $\omega_{\mathrm{0}} = 0$.

When the flux is swept through the avoided crossing, Landau-Zener tunneling can non-adiabatically flip the qubit state, at the probability $p_{\mathrm{LZ}}$. Outside the avoided crossing, the ground and excited states acquire a dynamical phase $\varphi = \pm \puoli \int  \omega (t) dt$ during the sweep. The phase is also contributed by the Stokes phase $\varphi_S$ acquired during the LZ event, given as $\varphi_S = 0$ (or $\varphi_S = \pi/4$) in the slow $p_{\mathrm{LZ}} \approx 0$ (or fast $p_{\mathrm{LZ}} \approx 1$) limit. If the sweep is repeated back and forth across the avoided crossing, the system acquires the dynamical phases $\varphi_{1,2}$ on either side. The phases can interfere constructively or destructively, resulting in oscillations of the qubit population as a function of the sweep parameters. The conditions of constructive interference, leading to enhanced population of the excited state, are then given by \cite{Ashhab2010LZ,Tuorila_2013,Silveri2017QuantumModulation}
\begin{subequations}
\begin{align}
\varphi_2 - \varphi_1 & = l \pi \label{res1}\\ 
\varphi_2 + \varphi_S & = m \pi \label{res2}
\end{align}
\end{subequations}
with integer $l, m$. Notice the arbitrary assignment of either $\varphi_1$ or $\varphi_2$ in \eref{res2}.
 
Now, let us consider our system that consists of a two-level system coupled to multiple bosonic fields, and how it can be understood as an extension of the two-state LZ problem. The system is described by the  multimode Jaynes-Cummings (MJC) model as
\begin{equation}
    H_\text{MJC}=\frac{\omega_{0}}{2}\sigma_{z}+\sum_{i}\omega_{m}^{(i)}a_{i}^{\dagger}a_{i}+\sum_{i}g^{(i)}_{m}(a_{i}\sigma_{+}+a_{i}^{\dagger}\sigma_{-})\,,
    \label{eq:H}
\end{equation}
where $\sigma_{z}$, $\sigma_{+}$ and $\sigma_{-}$ represent the standard qubit operators, and $a_{i}$ ($a_{i}^{\dagger}$) is the annihilation (creation) phonon operator for mode $i$ with frequency $\omega_{m}^{i}$. We study the behaviour of the system when the qubit is driven with both transverse (excitation) and longitudinal (frequency modulation)  classical fields: $H_{x}(t)=\Omega\cos{(\omega_\text{ext}t)}\sigma_{x}$ and $H_{z}(t)=\frac{A}{2}\cos{(\omega_\text{rf}t)}\sigma_{z}$, respectively. Here $\Omega$ is the excitation amplitude and $\omega_{\mathrm{ext}} \approx \omega_{0}$ is the excitation frequency. The full Hamiltonian is then $H(t)=H_\text{MJC}+H_{x}(t)+H_{z}(t)$.

In the rotating frame defined by the excitation frequency, the Hamiltonian becomes
\begin{equation}
\begin{split}
    H & = \frac{\Delta(t)}{2} \sigma_{z} + \frac{\Omega}{2} \sigma_{x}  +\sum_{i}\Delta_{i}a_{i}^{\dagger}a_{i} \\
    & + \sum_{i}g^{(i)}_{m} \left(a_{i}\sigma_{+}+a_{i}^{\dagger}\sigma_{-} \right) \,,
    \label{eq:HLZ}
    \end{split}
\end{equation}
%
with the detunings $\Delta(t) = \Delta_0 +A\cos{(\omega_\text{rf}t)}$, $\Delta_0 = \omega_0 -\omega_\text{ext}$, and $\Delta_i = \omega^{(i)}_{m} -\omega_\text{ext}$. The first two terms in Eq.~(\ref{eq:HLZ}) describe the regular LZS interference problem. One also uses the term photon assisted LZS interference \cite{Nori2012phLZ,Li2013,Silveri2015StuckelbergModulation}, since the qubit extracts a photon from the excitation field such that its energy is redefined as $\omega_{0}-\omega_\text{ext}$.

In our current case, we are concerned of the effect of the last two terms in \eref{eq:HLZ} on the LZS problem. Taking $\Omega$ to be much smaller than the other energy scales, the situation becomes that pictured in \fref{fig:LZscheme}. It describes modulated coupled energy levels, but they are those of a qubit and an oscillator, for a given oscillator $i$. Moreover, the qubit exhibits a similar coupling to many nondegenerate oscillators. We mention here a related recent experimental work \cite{Naik2017RandomElectrodynamics} that used a system of a qubit coupled to many electromagnetic cavities and showed stimulated vacuum Rabi oscillations, but did not treat the LZS limit.

On top of the picture of LZS modulation, the system allows for an interpretation in terms of multiphoton transitions \cite{supplement}, that manifest themselves as the appearance of sidebands in the spectrum \cite{Beaudoin2012First-orderModulation}. One obtains a time-independent effective Hamiltonian:
\begin{equation}
\begin{split}
        H_\text{eff}^{(n,k)}  & =  \frac{\Delta_0 + n\omega_\text{rf}}{2}\sigma_{z}+\sum_{i}\left(\Delta_i + k\omega_\text{rf} \right) a_{i}^{\dagger}a_{i}  \\
    & +\sum_{i}g^{(i)}_{m}J_{n-k}\left(\frac{A}{\omega_\text{rf}}\right) \left(\sigma_{+}a_{i}+\sigma_{-}a_{i}^{\dagger} \right) \\
    & + \frac{\Omega}{2}J_{n}\left(\frac{A}{\omega_\text{rf}}\right)\sigma_{x} \,.
\end{split}
\label{eq:Hmultiphot}
\end{equation}
$H_\text{eff}^{(n,k)}$ describes the interaction between the $n^\mathrm{th}$-order sideband of the qubit and k$^\textit{th}$-order sidebands of all the mechanical modes. The $n^\mathrm{th}$-order sideband of the qubit interacts with the $k^\mathrm{th}$-order sideband of a mechanical mode $i$ with coupling strength $g_\text{eff}^{n-k}=g^{(i)}_{m}J_{n-k}\left( \frac{A}{\omega_\text{rf}}\right)$, where $J_{j}$ are the Bessel functions of the first kind and order $j$. In other words, the qubit and one of the detuned mechanical modes take photons from the longitudinal field such that they become resonant and thus interact with each other at a rate $g_\text{eff}^{n-k}$.

To simulate the experimental results, we use \eref{eq:Hmultiphot} and determine the qubit population at the steady state by solving the Lindblad master equation including qubit losses, and limit the Hamiltonian to the first excitation manifold where only the zero and the one-phonon Fock states are considered \cite{Moores2018}. This is well justified because the mechanical resonator is already in the ground state and the qubit excitation amplitude is small in comparison with its linewidth.

In the experiment, our device consists of a flux-tunable transmon qubit coupled to an acoustic resonator (HBAR) whose piezoelectric effect enables a strong interaction between the electric fields of the qubit and the acoustic waves of the resonator. In contrast to other work using AlN \cite{SchoelkopfHBAR2017,Chu2018}, our sample is fabricated on an epitaxial Gallium Nitride (GaN) - coated Si substrate. The piezoelectric GaN film has been etched away everywhere else except directly under the qubit (\fref{fig:qubit}(a)). 
As seen in \fref{fig:qubit}(b-c), our transmon qubit has an asymmetric "pentamon" geometry with no parallel sides to greatly suppress lateral spurious modes of the acoustic resonator. The device is measured at the base temperature of a dilution refrigerator. At the ambient temperature inside the cryostat, both the qubit and the high GHz frequency mechanical modes reside naturally in their quantum ground state.

\begin{figure}[h]
  \begin{center}
    {\includegraphics[width=0.45\textwidth]{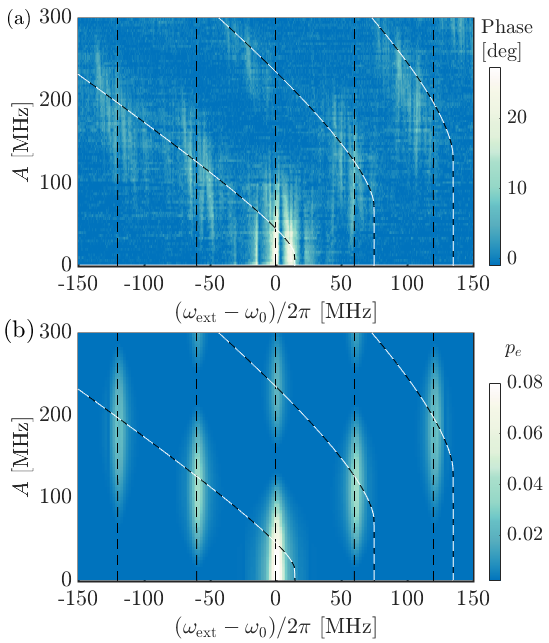}} 
    \caption{\emph{LZS dynamics in the multimode electromechanical system}.  (a) Experimental data depicting the qubit population when the slow modulation amplitude is varied. (b) Simulation of the qubit population \textit{without} presence of the acoustic modes in an otherwise similar situation. In both (a) and (b), the dashed black (black-white) lines display the respective LZS resonance conditions [\eref{res1}, \eref{res2}]. The modulation frequency is $\omega_{\mathrm{rf}}/2\pi = 60$ MHz.}
    \label{fig:LZamplitude}
 \end{center}
\end{figure}

The qubit-HBAR hybrid is coupled to a quarter-wavelength coplanar waveguide resonator that allows to interact with the system.
The qubit spectrum is shown in \fref{fig:qubit}(e). The qubit experiences avoided crossings spaced by the free spectral range $f_\text{FSR} = \frac{v}{2 T} \simeq 17.4~\text{MHz}$ of the acoustic modes. The latter is determined by the thickness of the substrate $T \simeq 270$ $\mu$m and the speed of sound $v \simeq 9400$ m/s. The interaction strength between the qubit and a single acoustic mode is $2 g_m\approx~$12 MHz interpreted from the vacuum Rabi splitting in \fref{fig:qubit}(d). With the total qubit linewidth $\gamma \approx 8$ MHz, the system is close to the strong coupling limit.

Next, we park the static DC flux at one spot on the qubit energy curve where the slope of the curve is close to linear. We apply the longitudinal modulation given by $H_z(t)$ on top of the static field to modulate the qubit energy around $\omega_{0}$. 
In  \fref{fig:LZamplitude} (a), we display the behavior of the qubit population when the longitudinal modulation amplitude is varied at a fixed modulation frequency. The qubit population is maximized around parameter regions satisfying both the interference conditions, Eqs.~(\ref{res1},\ref{res2}). The latter is clearly illustrated in \fref{fig:LZamplitude}(b), which is a simulation on a single qubit alone and hence describes the regular LZS situation. In the experimental data, however, the regions of constructive interference exhibit additional fine structure on top of the LZS interference pattern. We attribute the observed bending of the experimental data to the left in  \fref{fig:LZamplitude}(a) to a combination of curvature in the qubit frequency-flux relation and a flux drift during the measurement.

\begin{figure}[h]
  \begin{center}
    {\includegraphics[width=0.45\textwidth]{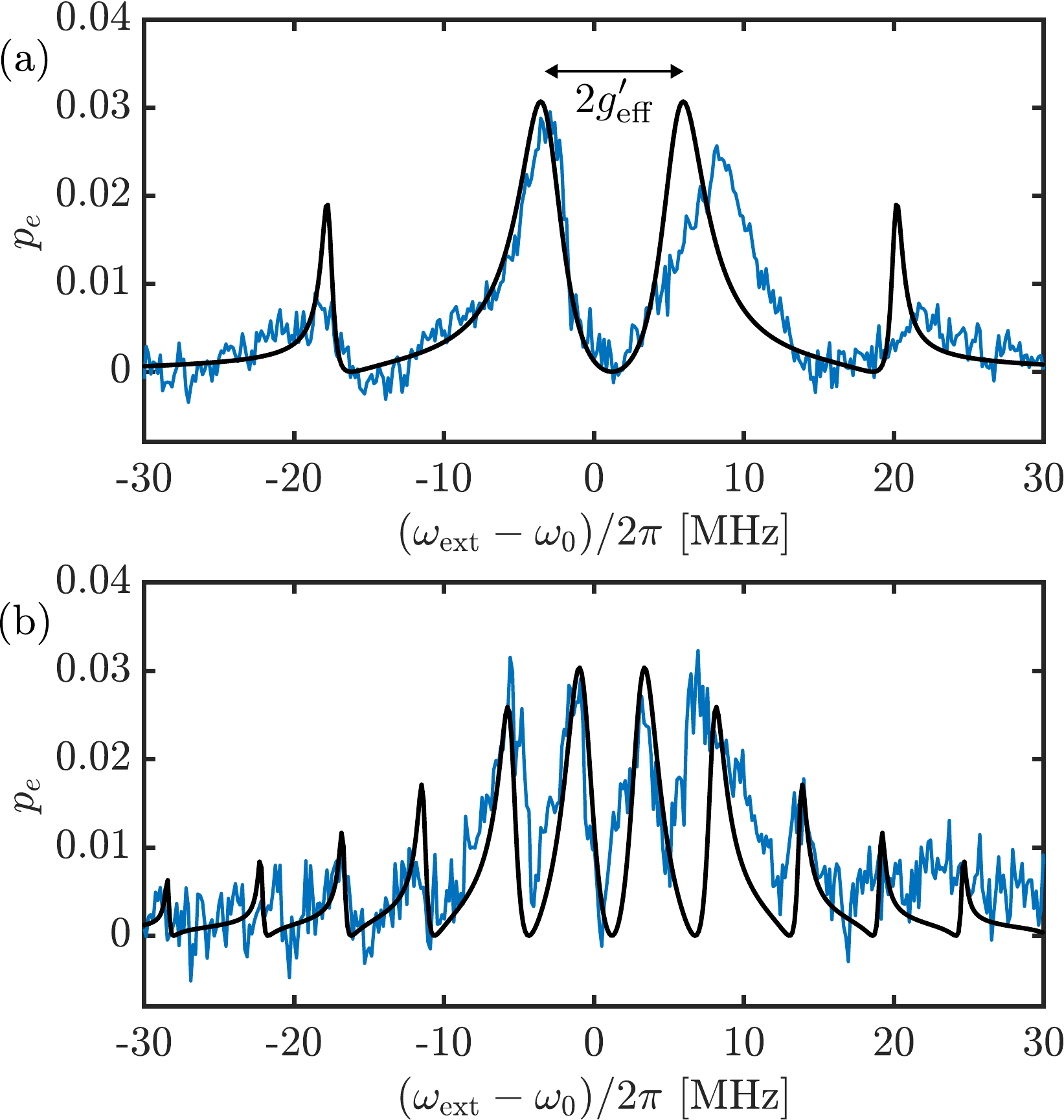}} 
    \caption{\textit{Resonances in the rotating frame}. (a) Excited state probability of the qubit under low-frequency  modulation with $\omega_{\mathrm{rf}}/2\pi = 139$ MHz. (b) With $\omega_{\mathrm{rf}}/2\pi = 148$ MHz. The black solid lines in both (a), (b) are numerical simulation of the master equation with $A = 210$ MHz, $\gamma/2\pi = 8$ MHz, $g_m =$ 5.5 MHz and $\Omega/2\pi =3$ MHz. We can infer that in (b), the sideband acoustic modes exhibit entanglement characterized by logarithmic negativity on the order 0.07.}
    \label{fig:linecuts}
 \end{center}
\end{figure}

In order to describe the additional resonances in \fref{fig:LZamplitude}(b), we adopt the multiphoton picture in \eref{eq:Hmultiphot}. The multiphoton transitions involve both the qubit, and each mechanical mode. For example, when the transverse driving field satisfies the condition $\Delta_i + k_i\omega_\text{rf} = \Delta_j + k_j\omega_\text{rf}$, two mechanical modes $i$ and $j$ become resonant. The case is extended to any number of modes. Moreover, if the effective qubit splitting $\puoli(\Delta_0 + n\omega_\text{rf})$ also satisfies the equality, the qubit is also on resonance with them. In \fref{fig:linecuts}(a) we display the latter situation. Three modes $i=307$, $j=323$ and $h=315$ form a tripartite resonance when $\omega_\text{rf} = 139$ MHz $=8\times f_{\mathrm{fsr}}$ with $k_i=1$, $k_j=-1$, $k_h=0$, and with the qubit at $\Delta_0\simeq 0$ and $n=0$. The effective vacuum Rabi splitting $\approx 11$ MHz is nearly as large as seen in the non-modulated case shown in \fref{fig:qubit}(d), although the simplest expectation yields $2 g_\text{eff}^{\pm 1} \simeq 6$ MHz. Instead, the vacuum Rabi splitting is that of co-resonant fourpartite ($N=4$ below) system formed by three oscillators and a qubit. In the present case, the couplings $g_\text{eff}^{\pm 1}$ and $g_\text{eff}^{0}$ are nearly equal, and the total coupling is $2 g_\text{eff}' \approx \sqrt{N-1}\times 2 g_\text{eff}^{\pm 1} \simeq 10.6$ MHz, in a good agreement with the measurement.

When the subsystems are brought off-resonant by detuning the modulation frequency as shown in \fref{fig:linecuts}(b), the system is understood as several detuned resonators that do not exhibit appreciable energy exchange.

\begin{figure}[h]
  \begin{center}
    {\includegraphics[width=0.48\textwidth]{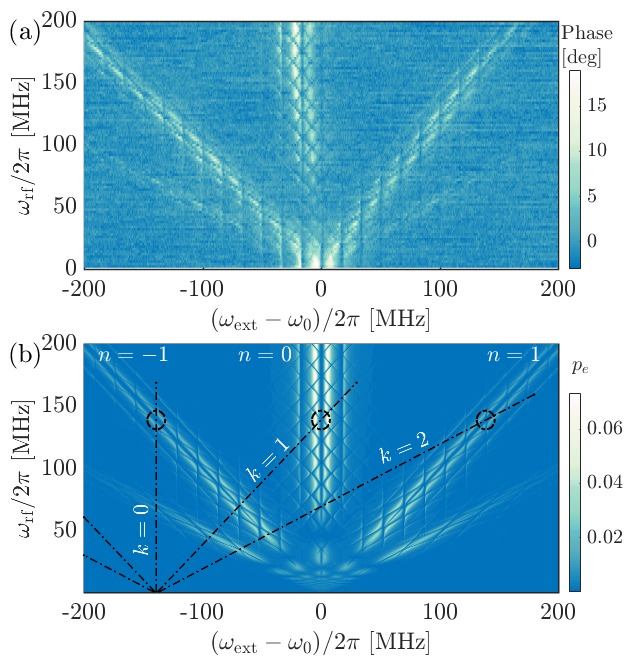}} 
    \caption{\textit{Multiphoton transitions in the multimode system.} (a) The measured phase shift shows how the qubit experiences resolved sidebands that are detuned from the original resonance by the frequency of the applied modulation. (b) Numerical simulation showing the qubit population, with parameters $A = 110$ MHz, $\gamma/2\pi = 8$ MHz, $g_m =$ 5.5 MHz and $\Omega/2\pi =3$ MHz. The qubit sidebands are marked with different $n$, and the mechanical sidebands by $k$ values.}
    \label{fig:qubitmodulation}
 \end{center}
\end{figure} 

The resonance conditions can be illustrated by plotting the qubit population as a function of two control parameters. In \fref{fig:qubitmodulation}(a) we can observe, first of all, the resolved sidebands in the spectrum at frequencies $\omega_\text{ext}=\omega_0\pm n \omega_\text{rf}$ $(n=0,1,2,...)$, see \eref{eq:Hmultiphot}. The interaction is mediated to multiple acoustic modes that exhibit sidebands as well. Each mechanical mode represents a starting point for a set of sideband transitions ($\omega \approx \omega_{m}^{(i)} \pm k\omega_\text{rf}$, $k=0,1,2,...$). They are easily identified in the measurement (\fref{fig:qubitmodulation}(a)) and in the corresponding simulation (\fref{fig:qubitmodulation}(b)). At the lowest frequencies below the bias-T cut-off, the modulation does not reach the qubit, and the measurement in this region is hence equivalent to a non-modulated system.

In the central band we see diagonal anticrossings separated by the free spectral range $f_\text{FSR} = 17.4~\text{MHz}$. For example, when the modulation frequency is 130-170 MHz we see the interaction of mechanical modes $\omega_m^{(i)}, i=315\pm8,\pm9,\pm10$ with the qubit, see \cite{supplement}. Therefore by selecting the frequency of the modulation to match  $\omega_0-\omega_m^{(i)}$, different acoustic modes can be brought into resonance with the qubit allowing addressing and hybridizing of different modes.

We have shown that a quantum electromechanical system under frequency modulation can be understood starting from Landau-Zener-St\"uckelberg interference. The modulation leads to the generation of sidebands, and the coupling rate can be controlled by the modulation field $\tfrac{A}{\omega_\text{rf}}$ in order to prepare, manipulate and measure each mechanical mode. The work enables to selectively configure mechanical modes at mismatched frequencies to interact with the qubit, giving promise for use in quantum information.

\begin{acknowledgments} This work was supported by the Academy of Finland (contracts 308290, 307757), by the European Research Council (contract 615755), by the Centre for Quantum Engineering at Aalto University, by The Finnish Foundation for Technology Promotion, and by the Wihuri Foundation. We acknowledge funding from the European Union's Horizon 2020 research and innovation program under grant agreement No.~732894 (FETPRO HOT). J.E.R.M. acknowledges support from UN-DIEB project HERMES 41611, and from COLCIENCIAS under project HERMES 31361 (code 110171249692) and ``Beca de Doctorados Nacionales de COLCIENCIAS 727''. We acknowledge the facilities and technical support of Otaniemi research infrastructure for Micro and Nanotechnologies (OtaNano). We would like to thank Herbert Vinck Posada for helpful discussions on the theoretical calculations.
\end{acknowledgments}


%

\end{document}